\documentclass[aps,prb,twocolumn,showpacs,groupedaddress]{revtex4-1}
\usepackage{amssymb}

\usepackage{graphicx}

\begin{document}

\title[]{Shubnikov-de Haas oscillations from topological surface states of metallic Bi$_2$Se$_{2.1}$Te$_{0.9}$}

\author{Keshav Shrestha$^1$, Vera Marinova$^2$, Bernd Lorenz$^1$, and Paul C. W. Chu$^{1,3}$}

\affiliation{$^1$ TCSUH and Department of Physics, University of Houston, 3201 Cullen Blvd., Houston, Texas 77204, USA}

\affiliation{$^2$ Institute of Optical Materials and Technology, Bulgarian Academy of Sciences, Acad. G. Bontchev Str. 109, Sofia 1113, Bulgaria}

\affiliation{$^3$ Lawrence Berkeley National Laboratory, 1 Cyclotron Road, Berkeley, California 94720, USA}

\begin{abstract}
We have studied the quantum oscillations in the conductivity of metallic, p-type Bi$_2$Se$_{2.1}$Te$_{0.9}$. The dependence of the oscillations on the angle of the magnetic field with the surface as well as the Berry phase determined from the Landau level fan plot indicate that the observed oscillations arise from surface carriers with the characteristic Dirac dispersion. Several quantities characterizing the surface conduction are calculated employing the Lifshitz-Kosevich theory. The low value of the Fermi energy with respect to the Dirac point is consistent with the metallic character of the bulk hole carriers. We conclude that, due to the peculiar shape of the valence band, the Shubnikov-de Haas oscillations of the bulk carriers are shifted to higher magnetic fields which allows for the detection of the quantum oscillations from the topological surface states at lower field.
\end{abstract}

\pacs{73.20.At,73.20.-r,73.25.+i}

\maketitle

Topological insulators have attracted increasing attention in recent years because of novel, macroscopic quantum phenomena arising from a nontrivial topology of their electronic states described by a set of wave functions which span the Hilbert space. The nontrivial topology demands the occurrence of gapless surface or interface states at the transition to a trivial (or ordinary) insulator, or to the vacuum. In the latter case, the surface of a topological insulator becomes conducting and the surface states are protected by symmetry, e.g. the time reversal symmetry in 3D topological insulators. The surface states of 3D topological insulators show the characteristic Dirac dispersion and the quasiparticles in this state are massless Dirac fermions where the spin is locked to the momentum forming a helical spin state.

The number of 3D topological insulators predicted theoretically and detected experimentally is steadily increasing. Several excellent reviews of topological insulator materials and phenomena have recently been published.\cite{hasan:10,qi:11,ando:13} Among others, bismuth chalcogenides (Bi$_2$Te$_3$, Bi$_2$Se$_3$, etc.), crystallizing in a tetradymite structure, have been identified as promising topological insulating materials.\cite{xia:09,chen:09,hsieh:09}

Experimentally, the topological nature of surface states in 3D topological insulators can be verified by angle-resolved photoemission spectroscopy (ARPES) revealing the typical Dirac cone feature in the excitation spectrum, as shown for example in Bi$_2$Se$_3$\cite{pan:11} and Bi$_2$Te$_3$.\cite{chen:09} Transport measurements in magnetic fields have been employed to prove the existence of conducting surface states by studying the characteristic quantum Shubnikov-de Haas (S-dH) oscillations of the longitudinal and transverse conductivities. However, bulk conduction can interfere with the surface transport if the Fermi energy is not well positioned in the gap of the bulk excitation spectrum.\cite{qu:10,analytis:10,eto:10,cao:13} Therefore, various attempts have been made to control the Fermi energy and reduce the bulk carrier number and conductivity resulting in a semiconducting (insulating) temperature dependence of the resistivity from bulk electronic states and extending the study of binary bismuth chalcogenides to the ternary compound, Bi$_2$Te$_2$Se.\cite{ren:10,taskin:11,xiong:12} The controlled removal of excess electrons was achieved by Sn doping and Bi excess in Bi$_2$Te$_2$Se.\cite{kushwaha:14}

Bi$_2$Se$_2$Te was theoretically predicted as a topological insulator\cite{wang:11} and recent experiments have shown the characteristic topological features in the band structure and in transport measurements\cite{miyamoto:12,bao:12} In these experimental studies, the Fermi energy was located in the bulk band gap, well above the Dirac point, implying negative charge carriers (electrons) confirmed by Hall measurements. In this study, we have observed quantum oscillations from the topological surface states of p-type single crystals of Bi$_2$Se$_{2.1}$Te$_{0.9}$ despite their bulk metallic temperature dependence. The magneto-transport data were analyzed and parameters of the topological states are determined.

Single crystals of Bi$_2$Se$_{2.1}$Te$_{0.9}$ were grown by a modified Bridgman technique. The starting materials with high purity, Bi  (99.9999\%), Se (99.9999\%), and Te (99.9999\%), were mixed according to the desired compositions in encapsulated quartz ampoules of 20 mm diameter. The mixtures were annealed at 875 $^\circ$C for 48 h in order to obtain a homogenized melt. Then the melt was first cooled to 670 $^\circ$C with cooling speed of 0.5 $^\circ$C/h. Finally the crystals were cooled to room temperature at a rate of 10 $^\circ$C/h.

Magneto-transport measurements have been conducted using the ac-transport option of the Physical Property Measurement System (PPMS, Quantum Design) in magnetic fields up to 7 Tesla. Six gold contacts were sputtered on a freshly cleaved crystal face for standard resistivity and Hall measurements. Platinum wires were attached using silver paint. The thermoelectric power was determined by a low-frequency (0.1 Hz) ac technique employing a two-heater method generating a sinusoidal temperature gradient with an amplitude of 0.25 K. The temperature dependence of the resistivity is metallic below room temperature, as shown in Fig. 1. Hall measurements reveal that the charge carriers are positive (holes), as shown in the lower-right inset to Fig. 1. The bulk carrier density estimated from Hall data at 5 K is n=2x10$^{18}$ cm$^{-3}$, consistent with the metallic character of the bulk resistivity. The p-type nature of the carriers is also confirmed by the positive and large thermoelectric power, shown in the upper-left inset to Fig. 1.

\begin{figure}
\begin{center}
\includegraphics[angle=0, width=2.5 in]{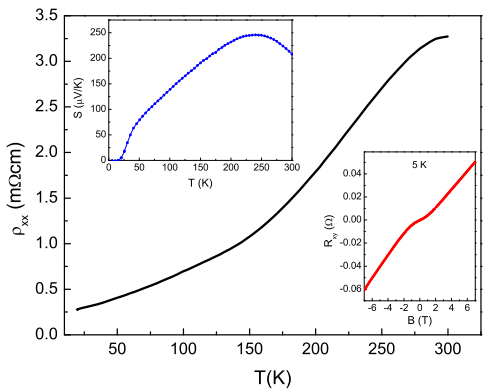}
\end{center}
\caption{(Color online) Temperature dependence of resistivity of a Bi$_2$Se$_{2.1}$Te$_{0.9}$ single crystal. The lower-right inset shows Hall data at 5 K. The upper-left inset displays the temperature dependence of the thermoelectric power.}
\end{figure}

The magnetoresistance $R_{xx}(B)$ of Bi$_2$Se$_{2.1}$Te$_{0.9}$ measured at 2 K is shown in Fig. 2. Shubnikov-de Haas oscillations are clearly visible at fields above 3 Tesla. Here the magnetic field was applied perpendicular to the large surface of the cleaved crystal of thickness of 100 $\mu m$. The derivative $dR_{xx}/dB$, also shown in Fig. 2 (right scale) reveals the existence of quantum oscillations more clearly, as expected from the cyclotron motion of the charge carriers in a perpendicular magnetic field. The oscillations are periodic with $1/B$ as is demonstrated in Fig. 3a by plotting $\Delta R_{xx}(B)$ as function of the inverse magnetic field ($\Delta R_{xx}$ was obtained by subtracting a smooth polynomial background function from $R_{xx}$). With increasing temperature, the amplitude of the oscillations decreases but the frequency does not change. The Fourier transform of the data, shown in Fig. 3b, shows the major sharp peak at a frequency of $F\approx$23 Tesla. At low temperatures, the Fourier transform displays a weak shoulder on the high-field side of the peak. This could indicate a small contribution from a higher frequency, as observed in other topological systems with complex Fermi surfaces.\cite{taskin:09,taskin:11,tu:14} However, the effect is relatively minor and disappears with increasing temperature.

\begin{figure}
\begin{center}
\includegraphics[angle=0, width=2.5 in]{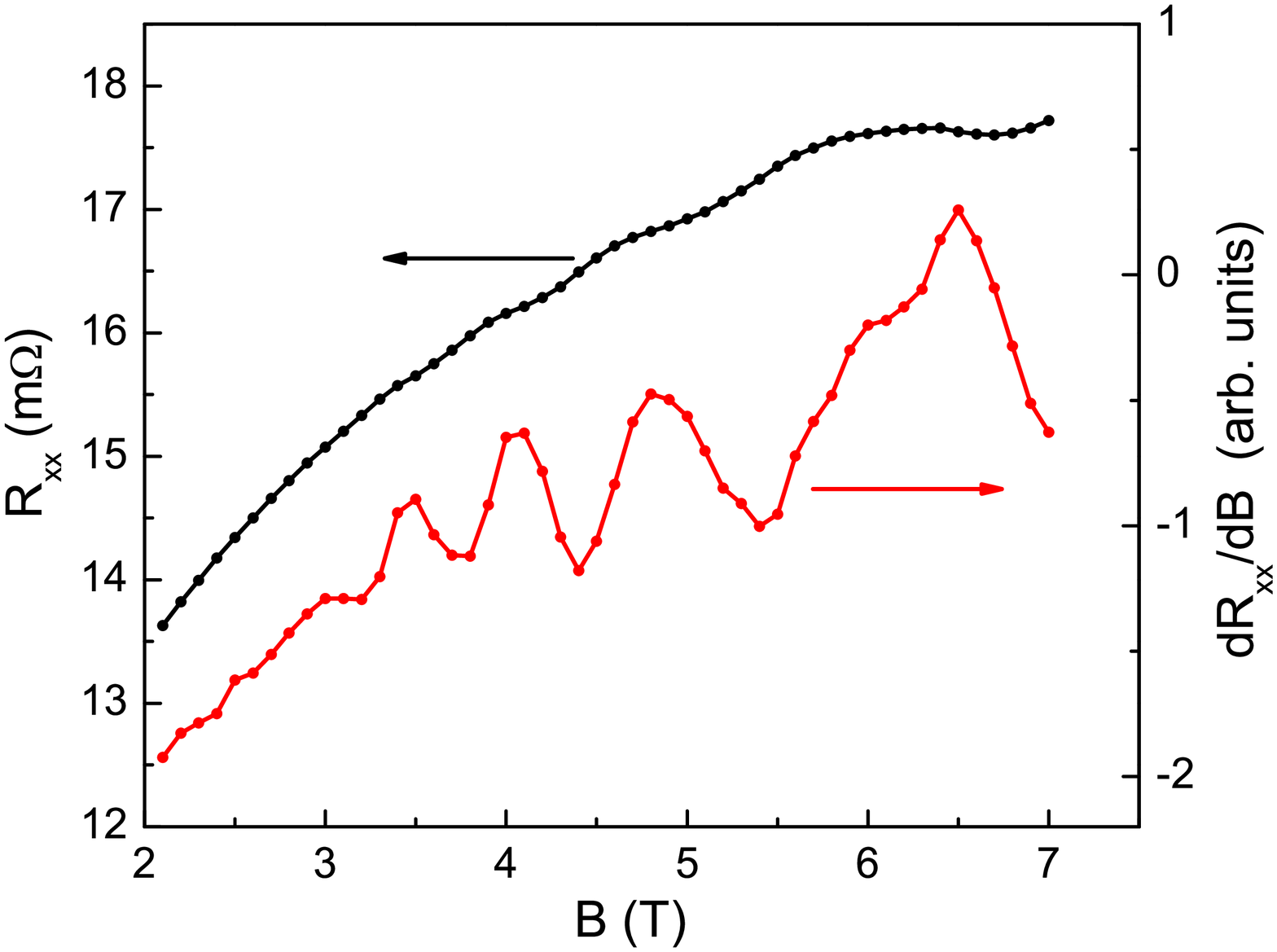}
\end{center}
\caption{(Color online) Magnetic field dependence of the resistance of Bi$_2$Se$_{2.1}$Te$_{0.9}$ (black, upper curve). The red (lower) curve shows the S-dH oscillations in the derivative, $dR_{xx}/dB$.}
\end{figure}

\begin{figure}
\begin{center}
\includegraphics[angle=0, width=3 in]{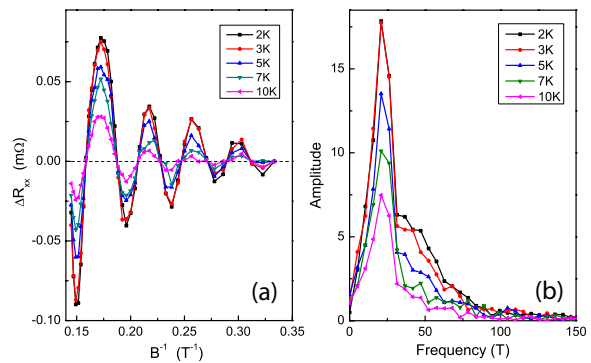}
\end{center}
\caption{(Color online) (a) Oscillatory part (S-dH) of the magnetoresistance, $\Delta R_{xx}(B)$, at different temperatures vs. inverse magnetic field. The field is perpendicular to the crystal's surface. (b) Fourier transform of the data from (a).}
\end{figure}

To identify the possible origin of the observed quantum oscillations, the angle dependence of the S-dH oscillations needs to be examined. Evaluating the derivative, $dR_{xx}/dB$, as function of the inverse applied magnetic field, $1/B$, for different angles $\Theta$ of the field with respect to the normal of the cleaved sample surface, the positions of the maxima and minima of the oscillation are found to change with $\Theta$, indicating the non-bulk nature of the oscillations. However, as demonstrated in Fig. 4a, the plot of $dR_{xx}/dB$ as function of $1/B_\perp=1/Bcos\Theta$ shows all maxima and minima lining up at the same position for all angles $\Theta$. The amplitude of the oscillations decreases quickly at higher angles and the S-dH oscillations cannot be resolved anymore. The angle dependence of one extremum is shown in Fig. 4b and it follows the expected $1/cos\Theta$ scaling for surface states. This is taken as evidence that the quantum oscillations arise from the topological surface states.\cite{qu:10}

\begin{figure}
\begin{center}
\includegraphics[angle=0, width=3 in]{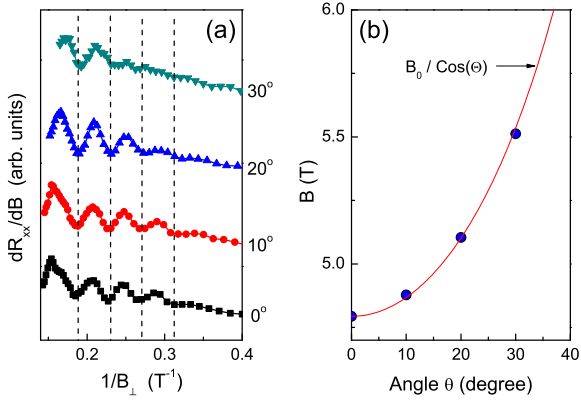}
\end{center}
\caption{(Color online) (a) $dR_{xx}/dB$ as function of the inverse perpendicular component, $1/B_\perp=1/Bcos\Theta$, demonstrating the scaling expected for surface conduction channels. (b) Plot of the field position of a maximum of $dR_{xx}/dB$ corresponding to n=5.5 in the Landau level fan diagram. The line shows the $1/cos(\Theta)$ scaling.}
\end{figure}

Further evidence of the origin of S-dH oscillations from topological surface states is the value of the Berry phase $\beta$ which can be extracted from the Landau level fan diagram.\cite{ando:13} Here the integer $n$, denoting the $n^{th}$ Landau level, is plotted as a function of the position of maxima and minima of the quantum oscillations, $1/B_{max/min}$. The value of $n_0$, obtained by a linear extrapolation of $1/B\rightarrow0$, defines the value of the Berry phase in units of 2$\pi$. $n_0=0.5$ ($\beta=\pi$) is expected for Dirac particles. As pointed out by Ando,\cite{ando:13} however, using resistivity $\rho_{xx}$ data can lead to deviations from the true value of $\beta$ and conductivity ($\sigma_{xx}$) data should be evaluated instead.

Fig. 5 shows the field dependence of the conductivity, calculated according to $\sigma_{xx}=\rho_{xx}/(\rho^2_{xx}+\rho^2_{xy})$ from $\rho_{xx}$ and $\rho_{xy}$ measured at 2 K. The S-dH oscillations are clearly visible in the derivatives of $\sigma_{xx}$ with respect to the field. The second derivative $d^2\sigma_{xx}/dB^2$ (upper inset to Fig. 5) was used to extract the values of $1/B_{max/min}$ used in the fan diagram, shown in the lower inset to Fig. 5. The positions of the extrema of $d^2\sigma_{xx}/dB^2$ are in perfect agreement with the extrema of $\Delta R_{xx}$ shown in Fig. 3a. The values from $\Delta R_{xx}$ are included as blue triangles and green diamonds in the fan diagram of Fig. 5. The linear extrapolation $1/B\rightarrow 0$ in the Landau level fan diagram yields $n_0=0.45 (\pm0.04)$, consistent with the Dirac nature of the particles, and a slope of $F$=23.3 Tesla, in good agreement with the characteristic frequency of the S-dH oscillations determined from Fig. 3b.

\begin{figure}
\begin{center}
\includegraphics[angle=0, width=3 in]{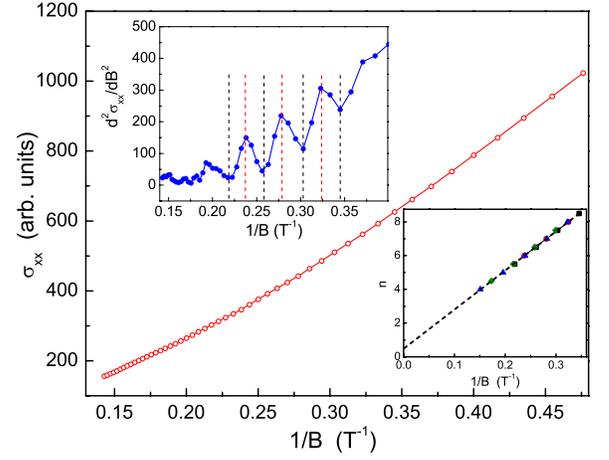}
\end{center}
\caption{(Color online) Field dependence of conductivity $\sigma_{xx}(B)$. Upper left inset: Second derivative, $d^2\sigma_{xx}/dB^2$ vs. $1/B$. The vertical dashed lines mark the positions of the maxima and minima of the quantum oscillations. Lower right inset: Landau level fan diagram with linear extrapolation (dashed line) to $1/B$=0.}
\end{figure}

The characteristic angle dependence and the value of the Berry phase show that the S-dH oscillations in the transport data are predominantly caused be topological surface states of Bi$_2$Se$_{2.1}$Te$_{0.9}$. This allows us to determine other characteristic parameters using the Lifshitz-Kosevich (LK) theory. The frequency of oscillations, $F=23.3$ Tesla, corresponds to a Fermi momentum $k_F=2.7\times10^6$ $cm^{-1}$ according to the Onsager relation, $F=\hbar/(2 e)k^2_F$. For a circular Fermi surface, this value of $k_F$ results in a surface carrier density of $n_{2D}=k^2_F/4\pi=5.8\times10^{11}$ $cm^{-2}$. The value of $k_F$ is slightly smaller than those obtained in other topological insulators with electron as well as hole carriers at the surface,\cite{analytis:10,qu:10,ren:10,taskin:11} indicating the closer proximity of the Fermi level to the Dirac point in our sample.

According to the LK theory, the temperature dependence of the amplitude of the S-dH oscillation is given by $R_T=\lambda(T/B)/sinh(\lambda(T/B)$, with $\lambda(T/B)=(2\pi^2k_BT/\Delta E_N(B)$ and the Landau level spacing $\Delta E_N(B)=\hbar eB/m_{cyc}$. $\Delta E_N(B)$ can be determined for different field values from $\Delta R_{xx}(T)$ as shown in Fig. 6 for $B=4.6$ Tesla. The lower left inset in Fig. 6 displays the linear dependence of $\Delta E_N$ on $B$. From the slope, the cyclotron mass is determined as $m_{cyc}=0.08m_0$, with $m_0$ the bare electron mass. With the linear dispersion relation for Dirac fermions, $v_F=\hbar k_F/m_{cyc}$, the Fermi velocity of the surface carriers is obtained as $v_F=3.9\times10^7$ $cm/s$.

Another factor in the LK theory is the Dingle factor, $e^{-\lambda_D}$ ($\lambda_D=2\pi^2k_BT_D/\hbar\omega_{cyc}$, $\omega_{cyc}$ is the cyclotron frequency), which accounts for the life time of the surface carries $\tau$ through the Dingle Temperature $T_D=\hbar/(2\pi k_B\tau)$. $T_D$ is determined, following the standard Dingle analysis, from the slope of the semi-logarithmic plot shown in the upper right inset to Fig. 6. With the estimated $T_D=12$ $K$, the surface carrier lifetime is $\tau=1.0\times10^{-13}s$, corresponding to a mean free path of $l_{2D}=v_F\tau=39$ $nm$ and a surface carrier mobility of $\mu_{2D}=(el_{2D})/(\hbar k_F)=2200$ $cm^2/(Vs)$. These values are comparable with other topological systems.\cite{xiong:12,taskin:12}

\begin{figure}
\begin{center}
\includegraphics[angle=0, width=3 in]{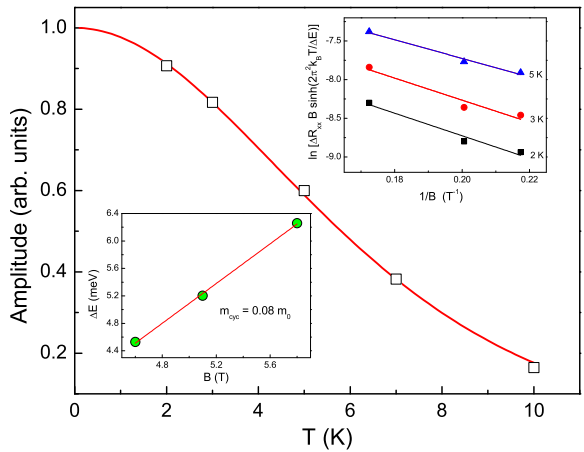}
\end{center}
\caption{(Color online) Temperature dependence of the amplitude of the S-dH oscillation ($\Delta R_{xx}$) at 4.6 Tesla. The red line represents the fit to the equation for $R_T$. Lower left inset: $\Delta E$ vs. $B$. Upper right inset: Dingle plot used to determine the Dingle temperature $T_D$ and the carrier life time $\tau$.}
\end{figure}

The results discussed above show that S-dH oscillations from topological surface states can be observed in Bi$_2$Se$_{2.1}$Te$_{0.9}$ despite the metallic conductivity from bulk carriers, which deserves a more detailed discussion. The relatively small value of $k_F$ indicates that the Fermi energy in our sample is lower than in other Bi-Se-Te based compounds. The estimated Fermi energy $E_F=69$ $meV$ is significantly closer to the Dirac point than $E_F$-values found for example in Bi$_2$Te$_3$,\cite{qu:10} Bi$_2$Te$_2$Se,\cite{ren:10} or in Sb-doped Bi$_2$Se$_3$.\cite{analytis:10} With this low value, $E_F$ cuts through the maxima of the valence band, as sketched in Fig. 7. Note that the valence band has two maxima at a finite momentum whereas the conduction band shows its minimum at $k=0$.\cite{ren:10,qu:10}

There arises the question why the bulk states do not produce S-dH oscillations within the field range of the current experiment. The peculiar shape of the valence band with the Fermi energy cutting through the two maxima as well as the Dirac cone requires a larger Fermi momentum, $k_F^{bulk}$, and a larger area of the Fermi surface to observe bulk S-dH oscillations. The corresponding oscillation frequency, $F=(\hbar/2e)k_F^2$ will be significantly higher. The magnetic field needed for the $n^{th}$ Landau level of the bulk carriers crossing the Fermi energy and impacting the conductivity is now much larger, dictated by the condition $F/B_n-\beta=n-1$. Therefore, the quantum oscillations of bulk carriers for low $n$ are shifted to higher fields, beyond the range of of the current measurements. Only bulk oscillations at larger Landau level indices could be observed at smaller fields, however, these oscillations are naturally attenuated, according to the LK-theory. This explains why in the present data (Figs. 2 to 4) the topological surface states dominate the oscillations of electrical transport properties. The observations discussed above open new possibilities to study topological effects in Bi-Te-Se type compounds when the Fermi energy is close to or even cutting though the valence band.

\begin{figure}
\begin{center}
\includegraphics[angle=0, width=2.5 in]{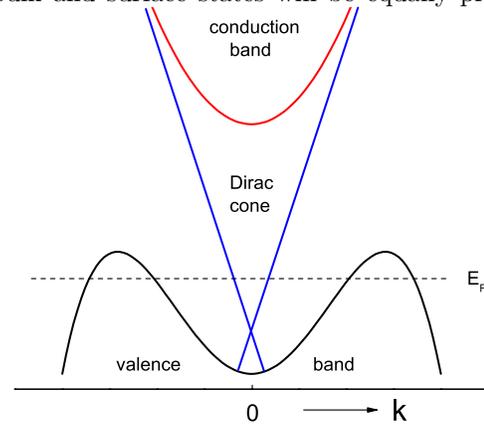}
\end{center}
\caption{(Color online) Schematic of the band structure of Bi$_2$Se$_{2.1}$Te$_{0.9}$. The Fermi energy is low enough to cut through the maxima of the valence band, resulting in bulk hole-like transport properties.}
\end{figure}

It should be noted that the above discussion only applies if the Fermi energy is low and close to the Dirac point and the valence band. For $E_F$ much higher, cutting through the bottom of the conduction band, the above argument is not valid. Since the conduction band has its minimum at $k=0$, the related Fermi momentum of the bulk carriers is of the same magnitude as that of the Dirac states. In this case, the bulk transport is electron-like (metallic) and the S-dH oscillations from bulk and surface states will be equally present in the whole range of magnetic fields. The quantum oscillations from bulk states are frequently dominating.\cite{qu:10} In a recent report, however, it was discussed that in electron-like metallic Bi$_2$Se$_3$ flakes, with the Fermi energy in the conduction band, S-dH oscillations from the topological surface state could be resolved, although the reason has yet to be explored.\cite{petrushevsky:12}

\textit{Note added:} After completion of this work we became aware of a preprint by Barua \textit{et al.}\cite{barua:14} reporting S-dH oscillations attributed to topological surface states in metallic, hole-like Bi$_2$Te$_3$. The observations in the low-field range up to 9 Tesla are consistent with our conclusions and they could be explained by the structure of the valence band, as discussed above.

\begin{acknowledgments}
This work is supported in part by the US Air Force Office of Scientific Research, the Robert A. Welch Foundation (E-1297), the T.L.L. Temple Foundation, the J. J. and R. Moores Endowment, and the State of Texas through the TCSUH and at LBNL by the DoE.
\end{acknowledgments}


\begin{thebibliography}{10}%
\makeatletter
\providecommand \@ifxundefined [1]{%
 \ifx #1\undefined \expandafter \@firstoftwo
 \else \expandafter \@secondoftwo
\fi
}%
\providecommand \@ifnum [1]{%
 \ifnum #1\expandafter \@firstoftwo
 \else \expandafter \@secondoftwo
\fi
}%
\providecommand \enquote [1]{``#1''}%
\providecommand \bibnamefont  [1]{#1}%
\providecommand \bibfnamefont [1]{#1}%
\providecommand \citenamefont [1]{#1}%
\providecommand\href[0]{\@sanitize\@href}%
\providecommand\@href[1]{\endgroup\@@startlink{#1}\endgroup\@@href}%
\providecommand\@@href[1]{#1\@@endlink}%
\providecommand \@sanitize [0]{\begingroup\catcode`\&12\catcode`\#12\relax}%
\@ifxundefined \pdfoutput {\@firstoftwo}{%
 \@ifnum{\z@=\pdfoutput}{\@firstoftwo}{\@secondoftwo}%
}{%
 \providecommand\@@startlink[1]{\leavevmode}%
 \providecommand\@@endlink[0]{}%
}{%
 \providecommand\@@startlink[1]{%
  \leavevmode
  \pdfstartlink
   attr{/Border[0 0 1 ]/H/I/C[0 1 1]}%
   user{/Subtype/Link/A<</Type/Action/S/URI/URI(#1)>>}%
  \relax
 }%
 \providecommand\@@endlink[0]{\pdfendlink}%
}%
\providecommand \url  [0]{\begingroup\@sanitize \@url }%
\providecommand \@url [1]{\endgroup\@href {#1}{\urlprefix}}%
\providecommand \urlprefix [0]{URL }%
\providecommand \Eprint[0]{\href }%
\@ifxundefined \urlstyle {%
  \providecommand \doi [1]{doi:\discretionary{}{}{}#1}%
}{%
  \providecommand \doi [0]{doi:\discretionary{}{}{}\begingroup
  \urlstyle{rm}\Url }%
}%
\providecommand \doibase [0]{http://dx.doi.org/}%
\providecommand \Doi[1]{\href{\doibase#1}}%
\providecommand \bibAnnote [3]{%
  \BibitemShut{#1}%
  \begin{quotation}\noindent
    \textsc{Key:}\ #2\\\textsc{Annotation:}\ #3%
  \end{quotation}%
}%
\providecommand \bibAnnoteFile [2]{%
  \IfFileExists{#2}{\bibAnnote {#1} {#2} {\input{#2}}}{}%
}%
\providecommand \typeout [0]{\immediate \write \m@ne }%
\providecommand \selectlanguage [0]{\@gobble}%
\providecommand \bibinfo [0]{\@secondoftwo}%
\providecommand \bibfield [0]{\@secondoftwo}%
\providecommand \translation [1]{[#1]}%
\providecommand \BibitemOpen[0]{}%
\providecommand \bibitemStop [0]{}%
\providecommand \bibitemNoStop [0]{.\EOS\space}%
\providecommand \EOS [0]{\spacefactor3000\relax}%
\providecommand \BibitemShut [1]{\csname bibitem#1\endcsname}%
\bibitem{hasan:10}%
  \BibitemOpen
  \bibfield{author}{%
  \bibinfo {author} {\bibfnamefont{M.~Z.}\ \bibnamefont{Hasan}}\ and\ \bibinfo
  {author} {\bibfnamefont{C.~L.}\ \bibnamefont{Cane}},\ }%
  \bibfield{journal}{%
  \bibinfo {journal} {Rev. Mod. Phys.}\ }%
  \textbf{\bibinfo {volume} {82}},\ \bibinfo {pages} {3045} (\bibinfo {year}
  {2010})%
  \bibAnnoteFile{NoStop}{hasan:10}%
\bibitem{qi:11}%
  \BibitemOpen
  \bibfield{author}{%
  \bibinfo {author} {\bibfnamefont{X.-L.}\ \bibnamefont{Qi}},\ }%
  \bibfield{journal}{%
  \bibinfo {journal} {Rev. Mod. Phys.}\ }%
  \textbf{\bibinfo {volume} {83}},\ \bibinfo {pages} {1057} (\bibinfo {year}
  {2011})%
  \bibAnnoteFile{NoStop}{qi:11}%
\bibitem{ando:13}%
  \BibitemOpen
  \bibfield{author}{%
  \bibinfo {author} {\bibfnamefont{Y.}~\bibnamefont{Ando}},\ }%
  \bibfield{journal}{%
  \bibinfo {journal} {J. Phys. Soc. Jpn.}\ }%
  \textbf{\bibinfo {volume} {82}},\ \bibinfo {pages} {102001} (\bibinfo {year}
  {2013})%
  \bibAnnoteFile{NoStop}{ando:13}%
\bibitem{xia:09}%
  \BibitemOpen
  \bibfield{author}{%
  \bibinfo {author} {\bibfnamefont{Y.}~\bibnamefont{Xia}}, \bibinfo {author}
  {\bibfnamefont{D.}~\bibnamefont{Qian}}, \bibinfo {author}
  {\bibfnamefont{D.}~\bibnamefont{Hsieh}}, \bibinfo {author}
  {\bibfnamefont{L.}~\bibnamefont{Wray}}, \bibinfo {author}
  {\bibfnamefont{A.}~\bibnamefont{Pal}}, \bibinfo {author}
  {\bibfnamefont{H.}~\bibnamefont{Lin}}, \bibinfo {author}
  {\bibfnamefont{A.}~\bibnamefont{Bansil}}, \bibinfo {author}
  {\bibfnamefont{D.}~\bibnamefont{Grauer}}, \bibinfo {author}
  {\bibfnamefont{Y.~S.}\ \bibnamefont{Hor}}, \bibinfo {author}
  {\bibfnamefont{R.~J.}\ \bibnamefont{Cava}},\ and\ \bibinfo {author}
  {\bibfnamefont{M.~Z.}\ \bibnamefont{Hasan}},\ }%
  \bibfield{journal}{%
  \bibinfo {journal} {Nat. Phys.}\ }%
  \textbf{\bibinfo {volume} {5}},\ \bibinfo {pages} {398} (\bibinfo {year}
  {2009})%
  \bibAnnoteFile{NoStop}{xia:09}%
\bibitem{chen:09}%
  \BibitemOpen
  \bibfield{author}{%
  \bibinfo {author} {\bibfnamefont{Y.~L.}\ \bibnamefont{Chen}}, \bibinfo
  {author} {\bibfnamefont{J.~G.}\ \bibnamefont{Analytis}}, \bibinfo {author}
  {\bibfnamefont{J.-H.}\ \bibnamefont{Chu}}, \bibinfo {author}
  {\bibfnamefont{Z.~K.}\ \bibnamefont{Liu}}, \bibinfo {author}
  {\bibfnamefont{S.-K.}\ \bibnamefont{Mo}}, \bibinfo {author}
  {\bibfnamefont{X.~L.}\ \bibnamefont{Qi}}, \bibinfo {author}
  {\bibfnamefont{H.~J.}\ \bibnamefont{Zhang}}, \bibinfo {author}
  {\bibfnamefont{D.~H.}\ \bibnamefont{Lu}}, \bibinfo {author}
  {\bibfnamefont{X.}~\bibnamefont{Dai}}, \bibinfo {author}
  {\bibfnamefont{Z.}~\bibnamefont{Fang}}, \bibinfo {author}
  {\bibfnamefont{S.~C.}\ \bibnamefont{Zhang}}, \bibinfo {author}
  {\bibfnamefont{I.~R.}\ \bibnamefont{Fisher}}, \bibinfo {author}
  {\bibfnamefont{Z.}~\bibnamefont{Hussain}},\ and\ \bibinfo {author}
  {\bibfnamefont{Z.-X.}\ \bibnamefont{Shen}},\ }%
  \bibfield{journal}{%
  \bibinfo {journal} {Science}\ }%
  \textbf{\bibinfo {volume} {325}},\ \bibinfo {pages} {178} (\bibinfo {year}
  {2009})%
  \bibAnnoteFile{NoStop}{chen:09}%
\bibitem{hsieh:09}%
  \BibitemOpen
  \bibfield{author}{%
  \bibinfo {author} {\bibfnamefont{D.}~\bibnamefont{Hsieh}}, \bibinfo {author}
  {\bibfnamefont{Y.}~\bibnamefont{Xia}}, \bibinfo {author}
  {\bibfnamefont{D.}~\bibnamefont{Qian}}, \bibinfo {author}
  {\bibfnamefont{L.}~\bibnamefont{Wray}}, \bibinfo {author}
  {\bibfnamefont{F.}~\bibnamefont{Meier}}, \bibinfo {author}
  {\bibfnamefont{J.~H.}\ \bibnamefont{Dil}}, \bibinfo {author}
  {\bibfnamefont{J.}~\bibnamefont{Osterwalder}}, \bibinfo {author}
  {\bibfnamefont{L.}~\bibnamefont{Patthey}}, \bibinfo {author}
  {\bibfnamefont{A.~V.}\ \bibnamefont{Fedorov}}, \bibinfo {author}
  {\bibfnamefont{H.}~\bibnamefont{Lin}}, \bibinfo {author}
  {\bibfnamefont{A.}~\bibnamefont{Bansil}}, \bibinfo {author}
  {\bibfnamefont{D.}~\bibnamefont{Grauer}}, \bibinfo {author}
  {\bibfnamefont{Y.~S.}\ \bibnamefont{Hor}}, \bibinfo {author}
  {\bibfnamefont{R.~J.}\ \bibnamefont{Cava}},\ and\ \bibinfo {author}
  {\bibfnamefont{M.~Z.}\ \bibnamefont{Hasan}},\ }%
  \bibfield{journal}{%
  \bibinfo {journal} {Phys. Rev. Lett.}\ }%
  \textbf{\bibinfo {volume} {103}},\ \bibinfo {pages} {146401} (\bibinfo {year}
  {2009})%
  \bibAnnoteFile{NoStop}{hsieh:09}%
\bibitem{pan:11}%
  \BibitemOpen
  \bibfield{author}{%
  \bibinfo {author} {\bibfnamefont{Z.-H.}\ \bibnamefont{Pan}}, \bibinfo
  {author} {\bibfnamefont{E.}~\bibnamefont{Vescovo}}, \bibinfo {author}
  {\bibfnamefont{A.~V.}\ \bibnamefont{Fedorov}}, \bibinfo {author}
  {\bibfnamefont{D.}~\bibnamefont{Gardner}}, \bibinfo {author}
  {\bibfnamefont{Y.~S.}\ \bibnamefont{Lee}}, \bibinfo {author}
  {\bibfnamefont{S.}~\bibnamefont{Chu}}, \bibinfo {author}
  {\bibfnamefont{G.~D.}\ \bibnamefont{Gu}},\ and\ \bibinfo {author}
  {\bibfnamefont{T.}~\bibnamefont{Valla}},\ }%
  \bibfield{journal}{%
  \bibinfo {journal} {Phys. Rev. Lett.}\ }%
  \textbf{\bibinfo {volume} {106}},\ \bibinfo {pages} {257004} (\bibinfo {year}
  {2011})%
  \bibAnnoteFile{NoStop}{pan:11}%
\bibitem{qu:10}%
  \BibitemOpen
  \bibfield{author}{%
  \bibinfo {author} {\bibfnamefont{D.-X.}\ \bibnamefont{Qu}}, \bibinfo {author}
  {\bibfnamefont{Y.~S.}\ \bibnamefont{Hor}}, \bibinfo {author}
  {\bibfnamefont{J.}~\bibnamefont{Xiong}}, \bibinfo {author}
  {\bibfnamefont{R.~J.}\ \bibnamefont{Cava}},\ and\ \bibinfo {author}
  {\bibfnamefont{N.~P.}\ \bibnamefont{Ong}},\ }%
  \bibfield{journal}{%
  \bibinfo {journal} {Science}\ }%
  \textbf{\bibinfo {volume} {329}},\ \bibinfo {pages} {821} (\bibinfo {year}
  {2010})%
  \bibAnnoteFile{NoStop}{qu:10}%
\bibitem{analytis:10}%
  \BibitemOpen
  \bibfield{author}{%
  \bibinfo {author} {\bibfnamefont{J.~G.}\ \bibnamefont{Analytis}}, \bibinfo
  {author} {\bibfnamefont{R.~D.}\ \bibnamefont{McDonald}}, \bibinfo {author}
  {\bibfnamefont{S.~C.}\ \bibnamefont{Riggs}}, \bibinfo {author}
  {\bibfnamefont{J.-H.}\ \bibnamefont{Chu}}, \bibinfo {author}
  {\bibfnamefont{G.~S.}\ \bibnamefont{Boebinger}},\ and\ \bibinfo {author}
  {\bibfnamefont{I.~R.}\ \bibnamefont{Fisher}},\ }%
  \bibfield{journal}{%
  \bibinfo {journal} {Nature Phys.}\ }%
  \textbf{\bibinfo {volume} {6}},\ \bibinfo {pages} {960} (\bibinfo {year}
  {2010})%
  \bibAnnoteFile{NoStop}{analytis:10}%
\bibitem{eto:10}%
  \BibitemOpen
  \bibfield{author}{%
  \bibinfo {author} {\bibfnamefont{K.}~\bibnamefont{Eto}}, \bibinfo {author}
  {\bibfnamefont{Z.}~\bibnamefont{Ren}}, \bibinfo {author}
  {\bibfnamefont{A.~A.}\ \bibnamefont{Taskin}}, \bibinfo {author}
  {\bibfnamefont{K.}~\bibnamefont{Segawa}},\ and\ \bibinfo {author}
  {\bibfnamefont{Y.}~\bibnamefont{Ando}},\ }%
  \bibfield{journal}{%
  \bibinfo {journal} {Phys. Rev. B}\ }%
  \textbf{\bibinfo {volume} {81}},\ \bibinfo {pages} {195309} (\bibinfo {year}
  {2010})%
  \bibAnnoteFile{NoStop}{eto:10}%
\bibitem{cao:13}%
  \BibitemOpen
  \bibfield{author}{%
  \bibinfo {author} {\bibfnamefont{H.}~\bibnamefont{Cao}}, \bibinfo {author}
  {\bibfnamefont{S.}~\bibnamefont{Xu}}, \bibinfo {author}
  {\bibfnamefont{I.}~\bibnamefont{Miotkowski}}, \bibinfo {author}
  {\bibfnamefont{J.}~\bibnamefont{Tian}}, \bibinfo {author}
  {\bibfnamefont{D.}~\bibnamefont{Pandey}}, \bibinfo {author}
  {\bibfnamefont{M.~Z.}\ \bibnamefont{Hasan}},\ and\ \bibinfo {author}
  {\bibfnamefont{Y.~P.}\ \bibnamefont{Chen}},\ }%
  \bibfield{journal}{%
  \bibinfo {journal} {Phys. Status Solidi RRL}\ }%
  \textbf{\bibinfo {volume} {7}},\ \bibinfo {pages} {133} (\bibinfo {year}
  {2013})%
  \bibAnnoteFile{NoStop}{cao:13}%
\bibitem{ren:10}%
  \BibitemOpen
  \bibfield{author}{%
  \bibinfo {author} {\bibfnamefont{Z.}~\bibnamefont{Ren}}, \bibinfo {author}
  {\bibfnamefont{A.~A.}\ \bibnamefont{Taskin}}, \bibinfo {author}
  {\bibfnamefont{S.}~\bibnamefont{Sasaki}}, \bibinfo {author}
  {\bibfnamefont{K.}~\bibnamefont{Segawa}},\ and\ \bibinfo {author}
  {\bibfnamefont{Y.}~\bibnamefont{Ando}},\ }%
  \bibfield{journal}{%
  \bibinfo {journal} {Phys. Rev. B}\ }%
  \textbf{\bibinfo {volume} {82}},\ \bibinfo {pages} {241306(R)} (\bibinfo
  {year} {2010})%
  \bibAnnoteFile{NoStop}{ren:10}%
\bibitem{taskin:11}%
  \BibitemOpen
  \bibfield{author}{%
  \bibinfo {author} {\bibfnamefont{A.~A.}\ \bibnamefont{Taskin}}, \bibinfo
  {author} {\bibfnamefont{Z.}~\bibnamefont{Ren}}, \bibinfo {author}
  {\bibfnamefont{S.}~\bibnamefont{Sasaki}}, \bibinfo {author}
  {\bibfnamefont{K.}~\bibnamefont{Segawa}},\ and\ \bibinfo {author}
  {\bibfnamefont{Y.}~\bibnamefont{Ando}},\ }%
  \bibfield{journal}{%
  \bibinfo {journal} {Phys. Rev. Lett.}\ }%
  \textbf{\bibinfo {volume} {107}},\ \bibinfo {pages} {016801} (\bibinfo {year}
  {2011})%
  \bibAnnoteFile{NoStop}{taskin:11}%
\bibitem{xiong:12}%
  \BibitemOpen
  \bibfield{author}{%
  \bibinfo {author} {\bibfnamefont{J.}~\bibnamefont{Xiong}}, \bibinfo {author}
  {\bibfnamefont{A.~C.}\ \bibnamefont{Petersen}}, \bibinfo {author}
  {\bibfnamefont{D.}~\bibnamefont{Qu}}, \bibinfo {author}
  {\bibfnamefont{Y.~S.}\ \bibnamefont{Hor}}, \bibinfo {author}
  {\bibfnamefont{R.~J.}\ \bibnamefont{Cava}},\ and\ \bibinfo {author}
  {\bibfnamefont{N.~P.}\ \bibnamefont{Ong}},\ }%
  \bibfield{journal}{%
  \bibinfo {journal} {Physica E}\ }%
  \textbf{\bibinfo {volume} {44}},\ \bibinfo {pages} {917} (\bibinfo {year}
  {2012})%
  \bibAnnoteFile{NoStop}{xiong:12}%
\bibitem{kushwaha:14}%
  \BibitemOpen
  \bibfield{author}{%
  \bibinfo {author} {\bibfnamefont{S.~K.}\ \bibnamefont{Kushwaha}}, \bibinfo
  {author} {\bibfnamefont{Q.~D.}\ \bibnamefont{Gibson}}, \bibinfo {author}
  {\bibfnamefont{J.}~\bibnamefont{Xiong}}, \bibinfo {author}
  {\bibfnamefont{I.}~\bibnamefont{Pletikosic}}, \bibinfo {author}
  {\bibfnamefont{A.~P.}\ \bibnamefont{Weber}}, \bibinfo {author}
  {\bibfnamefont{A.~V.}\ \bibnamefont{Fedorov}}, \bibinfo {author}
  {\bibfnamefont{N.~P.}\ \bibnamefont{Ong}}, \bibinfo {author}
  {\bibfnamefont{T.}~\bibnamefont{Valla}},\ and\ \bibinfo {author}
  {\bibfnamefont{R.~J.}\ \bibnamefont{Cava}},\ }%
  \bibinfo {note} {preprint: arXiv:cond-mat/1402.3870 (unpublished).}%
  \bibAnnoteFile{Stop}{kushwaha:14}%
\bibitem{wang:11}%
  \BibitemOpen
  \bibfield{author}{%
  \bibinfo {author} {\bibfnamefont{L.-L.}\ \bibnamefont{Wang}}\ and\ \bibinfo
  {author} {\bibfnamefont{D.~D.}\ \bibnamefont{Johnson}},\ }%
  \bibfield{journal}{%
  \bibinfo {journal} {Phys. Rev. B}\ }%
  \textbf{\bibinfo {volume} {83}},\ \bibinfo {pages} {241309(R)} (\bibinfo
  {year} {2011})%
  \bibAnnoteFile{NoStop}{wang:11}%
\bibitem{miyamoto:12}%
  \BibitemOpen
  \bibfield{author}{%
  \bibinfo {author} {\bibfnamefont{K.}~\bibnamefont{Miyamoto}}, \bibinfo
  {author} {\bibfnamefont{A.}~\bibnamefont{Kimura}}, \bibinfo {author}
  {\bibfnamefont{T.}~\bibnamefont{Okuda}}, \bibinfo {author}
  {\bibfnamefont{H.}~\bibnamefont{Miyahara}}, \bibinfo {author}
  {\bibfnamefont{K.}~\bibnamefont{Kuroda}}, \bibinfo {author}
  {\bibfnamefont{H.}~\bibnamefont{Namatame}}, \bibinfo {author}
  {\bibfnamefont{M.}~\bibnamefont{Taniguchi}}, \bibinfo {author}
  {\bibfnamefont{S.~V.}\ \bibnamefont{Eremeev}}, \bibinfo {author}
  {\bibfnamefont{T.~V.}\ \bibnamefont{Menshchikova}}, \bibinfo {author}
  {\bibfnamefont{E.~V.}\ \bibnamefont{Chulkov}}, \bibinfo {author}
  {\bibfnamefont{K.~A.}\ \bibnamefont{Kokh}},\ and\ \bibinfo {author}
  {\bibfnamefont{O.~E.}\ \bibnamefont{Tereshchenko}},\ }%
  \bibinfo {note} {preprint: arXiv:cond-mat/1203.4439 (unpublished).}%
  \bibAnnoteFile{Stop}{miyamoto:12}%
\bibitem{bao:12}%
  \BibitemOpen
  \bibfield{author}{%
  \bibinfo {author} {\bibfnamefont{L.}~\bibnamefont{Bao}}, \bibinfo {author}
  {\bibfnamefont{L.}~\bibnamefont{He}}, \bibinfo {author}
  {\bibfnamefont{N.}~\bibnamefont{Meyer}}, \bibinfo {author}
  {\bibfnamefont{X.}~\bibnamefont{Kou}}, \bibinfo {author}
  {\bibfnamefont{P.}~\bibnamefont{Zhang}}, \bibinfo {author}
  {\bibfnamefont{Z.-G.}\ \bibnamefont{Chen}}, \bibinfo {author}
  {\bibfnamefont{A.}~\bibnamefont{Fedorov}}, \bibinfo {author}
  {\bibfnamefont{J.}~\bibnamefont{Zou}}, \bibinfo {author}
  {\bibfnamefont{T.~M.}\ \bibnamefont{Riedemann}}, \bibinfo {author}
  {\bibfnamefont{T.~A.}\ \bibnamefont{Lograsso}}, \bibinfo {author}
  {\bibfnamefont{K.~L.}\ \bibnamefont{Wang}}, \bibinfo {author}
  {\bibfnamefont{G.}~\bibnamefont{Tuttle}},\ and\ \bibinfo {author}
  {\bibfnamefont{F.}~\bibnamefont{Xiu}},\ }%
  \bibfield{journal}{%
  \bibinfo {journal} {Scientific Reports}\ }%
  \textbf{\bibinfo {volume} {2}},\ \bibinfo {pages} {726} (\bibinfo {year}
  {2012})%
  \bibAnnoteFile{NoStop}{bao:12}%
\bibitem{taskin:09}%
  \BibitemOpen
  \bibfield{author}{%
  \bibinfo {author} {\bibfnamefont{A.~A.}\ \bibnamefont{Taskin}}\ and\ \bibinfo
  {author} {\bibfnamefont{Y.}~\bibnamefont{Ando}},\ }%
  \bibfield{journal}{%
  \bibinfo {journal} {Phys. Rev. B}\ }%
  \textbf{\bibinfo {volume} {80}},\ \bibinfo {pages} {085303} (\bibinfo {year}
  {2009})%
  \bibAnnoteFile{NoStop}{taskin:09}%
\bibitem{tu:14}%
  \BibitemOpen
  \bibfield{author}{%
  \bibinfo {author} {\bibfnamefont{N.~H.}\ \bibnamefont{Tu}}, \bibinfo {author}
  {\bibfnamefont{Y.}~\bibnamefont{Tanabe}}, \bibinfo {author}
  {\bibfnamefont{K.~K.}\ \bibnamefont{Huynh}}, \bibinfo {author}
  {\bibfnamefont{Y.}~\bibnamefont{Sato}}, \bibinfo {author}
  {\bibfnamefont{H.}~\bibnamefont{Oguro}}, \bibinfo {author}
  {\bibfnamefont{S.}~\bibnamefont{Heguri}}, \bibinfo {author}
  {\bibfnamefont{K.}~\bibnamefont{Tsuda}}, \bibinfo {author}
  {\bibfnamefont{M.}~\bibnamefont{Terauchi}}, \bibinfo {author}
  {\bibfnamefont{K.}~\bibnamefont{Watanabe}},\ and\ \bibinfo {author}
  {\bibfnamefont{K.}~\bibnamefont{Tanigaki}},\ }%
  \bibfield{journal}{%
  \bibinfo {journal} {Appl. Phys. Lett.}\ }%
  \textbf{\bibinfo {volume} {105}},\ \bibinfo {pages} {063104} (\bibinfo {year}
  {2014})%
  \bibAnnoteFile{NoStop}{tu:14}%
\bibitem{taskin:12}%
  \BibitemOpen
  \bibfield{author}{%
  \bibinfo {author} {\bibfnamefont{A.~A.}\ \bibnamefont{Taskin}}, \bibinfo
  {author} {\bibfnamefont{S.}~\bibnamefont{Sasaki}}, \bibinfo {author}
  {\bibfnamefont{K.}~\bibnamefont{Segawa}},\ and\ \bibinfo {author}
  {\bibfnamefont{Y.}~\bibnamefont{Ando}},\ }%
  \bibfield{journal}{%
  \bibinfo {journal} {Adv. Mater.}\ }%
  \textbf{\bibinfo {volume} {24}},\ \bibinfo {pages} {5581} (\bibinfo {year}
  {2012})%
  \bibAnnoteFile{NoStop}{taskin:12}%
\bibitem{petrushevsky:12}%
  \BibitemOpen
  \bibfield{author}{%
  \bibinfo {author} {\bibfnamefont{M.}~\bibnamefont{Petrushevsky}}, \bibinfo
  {author} {\bibfnamefont{E.}~\bibnamefont{Lahoud}}, \bibinfo {author}
  {\bibfnamefont{A.}~\bibnamefont{Ron}}, \bibinfo {author}
  {\bibfnamefont{E.}~\bibnamefont{Maniv}}, \bibinfo {author}
  {\bibfnamefont{I.}~\bibnamefont{Diamant}}, \bibinfo {author}
  {\bibfnamefont{I.}~\bibnamefont{Neder}}, \bibinfo {author}
  {\bibfnamefont{S.}~\bibnamefont{Wiedmann}}, \bibinfo {author}
  {\bibfnamefont{V.~K.}\ \bibnamefont{Guduru}}, \bibinfo {author}
  {\bibfnamefont{F.}~\bibnamefont{Chiappini}}, \bibinfo {author}
  {\bibfnamefont{U.}~\bibnamefont{Zeitler}}, \bibinfo {author}
  {\bibfnamefont{J.~C.}\ \bibnamefont{Maan}}, \bibinfo {author}
  {\bibfnamefont{K.}~\bibnamefont{Chashka}}, \bibinfo {author}
  {\bibfnamefont{A.}~\bibnamefont{Kanigel}},\ and\ \bibinfo {author}
  {\bibfnamefont{Y.}~\bibnamefont{Dagan}},\ }%
  \bibfield{journal}{%
  \bibinfo {journal} {Phys. Rev. B}\ }%
  \textbf{\bibinfo {volume} {86}},\ \bibinfo {pages} {045131} (\bibinfo {year}
  {2012})%
  \bibAnnoteFile{NoStop}{petrushevsky:12}%
\bibitem{barua:14}%
  \BibitemOpen
  \bibfield{author}{%
  \bibinfo {author} {\bibfnamefont{S.}~\bibnamefont{Barua}}, \bibinfo {author}
  {\bibfnamefont{K.~P.}\ \bibnamefont{Rajeev}},\ and\ \bibinfo {author}
  {\bibfnamefont{A.~K.}\ \bibnamefont{Gupta}},\ }%
  \bibinfo {note} {preprint: arXiv:cond-mat/1409.1356v1 (unpublished).}%
  \bibAnnoteFile{Stop}{barua:14}%
\end{thebibliography}

%

\end{document}